\documentclass[prb,showpacs,floatfix,footinbib,twocolumn]{revtex4}
\usepackage{graphicx}
\usepackage{dcolumn}
\usepackage{epsfig}
\usepackage{float}
\usepackage{bm}

\usepackage{bibentry}

\begin{document}

\bibliographystyle{apsrev}

\title{Optical spectroscopy study on the electronic structure of \boldmath Eu$%
_{1-x} $Ca$_{x}$B$_{6}$ \unboldmath}

\author{ Jungho Kim and S.-J. Oh}
\affiliation{School of Physics Center for Strongly Correlated
Materials Research, Seoul National University, Seoul 151-742,
Republic of Korea}
\author{Youngwoo Lee and E. J. Choi\footnote{To whom all correspondences should be
addressed: echoi@uos.ac.kr.}}
\affiliation{Department of Physics,
University of Seoul, Seoul 130-743, Republic of Korea}
\author{C. C. Homes}
\affiliation{Department of Physics, Brookhaven National Laboratory,
Upton, NY 11973-5000}
\author{Jong-Soo Rhyee and B. K. Cho}
\affiliation{Center for Frontier Materials and Department of
Materials Science and Engineering, \\ K-JIST, Gwangju 500-712,
Korea}

%
% The abstract
%
\begin{abstract}

The optical conductivity $\sigma _{1}(\omega )$ of
Eu$_{1-x}$Ca$_{x}$B$_{6}$ has been obtained from reflectivity and
ellipsometry measurements for series of compositions, $0\leq x\leq
1$. The interband part of $\sigma _{1}(\omega ) $ shifts
continuously to higher frequency as Ca-content $x$ increases. Also
the intraband spectral weight of $\sigma _{1}(\omega )$ decreases
rapidly and essentially vanishes for $x\geq x_{c}=0.35$. These
results show that the valence band and the conduction band of
Eu$_{1-x}$Ca$_{x}$B$_{6}$ move away from each other such that their
band overlap decreases with increasing Ca-substitution. As a result,
the electronic state evolves from the semimetallic structure of
EuB$_{6}$ to the insulating CaB$_{6}$ where the two bands are
separated to open a finite gap ($\simeq 0.25$~eV) at the X-point of
the Brillouin zone.

\end{abstract}

\pacs{71.20.-b,71.30.+h,78.20.-e}

\maketitle

%
% The main body of the text
%
The divalent hexaboride compounds RB$_{6}$ (R=Eu, Ca, Sr, etc) have drawn
much attention for the last decade due to their interesting electrical and
magnetic properties. These materials crystalize in a cubic structure with
boron octahedra in the center of the unit cell and the cations sitting on
the corners of the cube. In EuB$_{6},$ the Eu ions ($S=7/2$) exhibit a
ferromagnetic alignment at $T_{C}=15$~K.\cite{matthias,kasaya} This
transition accompanies a large drop of dc-resistivity.\cite{fisk} In
addition, the infrared reflectivity measurement showed an unusual shift of
the plasma frequency across the transition.\cite{degiorgi} The resistivity
and the plasma frequency changes are also induced by external magnetic
field. \cite{broderick}

The hexaboride CaB$_{6}$ is isovalent with EuB$_{6}$. However, the
low-temperature dc-resistivity exhibits a semiconducting temperature
dependence, in contrast with the metallic behavior of EuB$_{6}$. While the
Ca ion bears no magnetic moment, weak ferromagnetism was observed at high
temperature ($T_{C}\sim 600$~K) in the lightly La-doped Ca$_{1-x}$La$_{x}$B$%
_{6}$ as well as in the nominally stoichiometric CaB$_{6}$ compound.\cite
{young1} Various explanations of this effect such as the excitonic state
model\cite{zhitomirsky,balents,barzykin,murakami1,murakami2} and the dilute
electron gas model\cite{young1,ceperley,ortiz} have been proposed. The
effect is also attributed to extrinsic origin such as structural defect or
impurities.\cite{gavilano,matsubayashi1,matsubayashi2,young2}

As a prerequisite to the understanding of these unconventional phenomena,
the band structure of EuB$_{6}$ and CaB$_{6}$ were extensively studied both
theoretically and experimentally. de Haas-van Alphen (dHvA) and Shubnikov de
Hass (SdH) experiments show a semimetallic band structure in both materials,
i.e, both electron pocket and hole pocket exist on the Fermi surface.\cite
{goodrich,aronson,terashima,hall} On the other hand, angle resolved
photoemission spectroscopy (ARPES) and soft x-ray emission measurements
showed that in both compounds the bands are separated with a sizable gap of $%
\sim 1$~eV.\cite{denlinger} Theoretically, the early local-density
approximation (LDA) band calculation predicted that, in EuB$_{6}$ and CaB$%
_{6}$, conduction band (CB) overlaps with valence band (VB) at X point of
the Brillouin zone.\cite{massidda} LDA+U calculation result for EuB$_{6}$
agrees with this semimetallic band structure.\cite{kunes} However, for CaB$%
_{6}$, pseudo-potential GW \cite{tromp,rodriguez} and also WDA
calculation \cite{wu} claim an insulating state with a sizable band
gap of 0.8 eV. On the contrary, all electron GW calculation by Kino
$et$ $al.$ shows that CaB$_{6}$ does not have such a large
gap.\cite{kino} A full-potential LMTO calculation yields a moderate
gap of 0.3 eV.\cite{kotani} This lack of consistency among different
results suggests that the band structure depends much on the details
of the employed calculation methods.

Recently, Rhyee \textit{et al.} have prepared a series of Eu$_{1-x}$Ca$_{x}$B%
$_{6}$ where Eu is gradually replaced by the isovalent Ca.\cite
{rhyee2,rhyee3} As the Ca-content $x$ increases, the metallic dc-resistivity
$\rho (T)$ was found to increase and then crosses over to an insulating
behavior. Noting that no carrier doping is expected in this series of
isovalent samples, the evident changes of the electrical properties imply a
non-trivial effect of the Ca-substitution. Also, this system provides an
opportunity to understand, when combined with a spectroscopic measurement,
how the electronic structure changes along with the cation substitution.
This will provide useful clues about the band structure of the two parent
compounds EuB$_{6}$ and CaB$_{6}$.

%
% Experimental
%
In this study, we have performed a wide range ($20-5\,0,000$~cm$^{-1}$)
optical spectroscopic measurement of Eu$_{1-x}$Ca$_{x}$B$_{6}$ for a serious
of compositions $x$ which ranges from 0 to 1. The single-crystal samples
were synthesized by a boro-thermal method as described in detail elsewhere.
\cite{rhyee2, rhyee3} Boron powder of 99.9\% purity was used. To determine
the Eu content $x$, we measured the dc-magnetization $M(H)$ of each sample
until it saturates at high magnetic field H. The saturation value is
proportional to $x$, from which we find $x=0,0.13,0.25,0.35,0.54$ and 1.
Dc-resistivity and Hall coefficient of these samples were reported in the
earlier publication. \cite{rhyee3} For present optical study, crystals from
the same sample batches were used. The reflectance $R(\omega )$ at a
near-normal angle of incidence was measured at 300~K in the $20-5\,000$~cm$%
^{-1}$ and $5\,000-50\,000$~cm$^{-1}$ ranges using a Fourier transform
spectrometer with an \textit{in situ} overcoating technique,\cite{homes93}
and a grating spectrometer with a V-W method, respectively. The crystals
were wedged by 2$%
%TCIMACRO{\UNICODE[m]{0xb0}}%
%BeginExpansion
{{}^\circ}%
%EndExpansion
$ to avoid internal interference effect.

%
% Figure 1
%
\begin{figure}[t]
\vspace*{-0.2cm}\centerline{%
\includegraphics[width=2.4in,angle=-90]{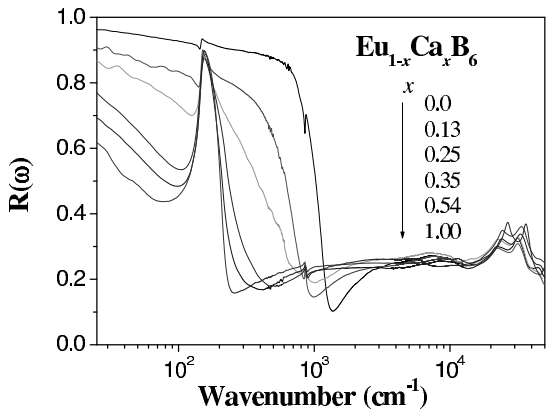}}
%\centerline{\includegraphics[width=4.5in,angle=-90]{fig1.eps}}%
\vspace*{0.0cm}%
%\centering\epsfig{file=fig1.eps,width=5cm,angle=-90}
%
\caption{Reflectivity spectra of Eu$_{1-x}$Ca$_{x}$B$_{6}$ for
various values of $x$ from $x=0$ (EuB$_{6}$) and $x=1$ (CaB$_{6}$)
taken at room temperature.} \label{fig:reflec}
\end{figure}

Figure~\ref{fig:reflec} shows the reflectivity of Eu$_{1-x}$Ca$_{x}$B$_{6}$
over the wide range of frequency. In the low energy infrared region, the
spectra exhibit a large change with $x$. For $x$=0 (EuB$_{6}$), R$(\omega )$
is high and shows a plasma edge at $\omega \sim $ 1,200 cm$^{-1}$, which is
similar to the earlier observation by Degiorgi $et$ $al.$. \cite{degiorgi}
As $x$ increases, the reflectivity level decreases and the plasma edge
shifts continuously toward lower frequency. For $x\geq $ 0.35, this metallic
feature of R$(\omega )$ is significantly suppressed and the phonon peak at
150 cm$^{-1}$ becomes prominent. The absorption peaks at high energy ($%
\omega >10^{4}$ cm$^{-1}$) correspond to interband excitations.

The real part of the optical conductivity $\sigma _{1}(\omega )$ has been
determined from a Kramers-Kronig analysis of the measured reflectance, for
which extrapolations for $\omega \rightarrow 0$ and $\infty $ must be
supplied. For $\omega \rightarrow 0$, the Hagen-Rubens (HR) extrapolation
was employed. Above the highest measured frequency, a free-electron
approximation $R(\omega )\propto \omega ^{-4}$ was assumed. For $1.5\leq
\omega \leq 5.5$ eV, an ellipsometer (Sopra GES5) was used to directly
determine $\sigma _{1}(\omega )$. The result showed a good agreement with
the reflectance measured in the same region.

%
% Figure 2
%
\begin{figure}[t]
%
%
% manuscript
%
\vspace*{0.4cm}\centerline{\includegraphics[width=2.6in]{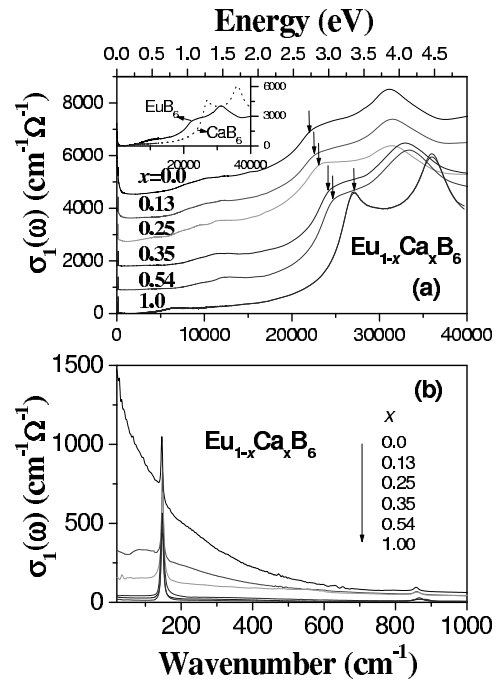}}
%\centerline{\includegraphics[width=4.5in]{fig2.eps}}%
\vspace*{-0.5cm}
\caption{Optical conductivity $\protect\sigma _{1}(\protect\omega )$ of Eu$%
_{1-x}$Ca$_{x}$B$_{6}$ for: (a) wide-range of frequency up to 5 eV;
(b)low-frequency infrared region. In (a), the curves are uniformly
displaced
along the vertical axis to avoid heavy overlap among them. Inset shows $%
\protect\sigma _{1}(\protect\omega )$ of EuB$_{6}$ and CaB$_{6}$
without the vertical shift.} \label{fig:sigma}
\end{figure}

Figure~\ref{fig:sigma}(a) displays the overall structure of $\sigma
_{1}(\omega )$ on a linear scale. To allow for a clearer presentation, the
curves are displaced uniformly along the vertical axis. The conductivity $%
\sigma _{1}(\omega )$ consists of a gradual rise up to $\simeq 2.5$~eV and
two eminent peaks at 2.7 and 3.7~eV, respectively. Also a weak absorption
appears at about 1.5~eV. Note that the 2.7~eV peak shifts to higher energy
as $x$ increases, as indicated by the arrows. A similar shift occurs with
the 3.7~eV peak as well. The inset displays $\sigma _{1}(\omega )$ of $x=0$
and $1$ without the vertical displacement. Note that the shift occurs over
the wide energy range, including the region of the gradual rise.

We assign the observed $\sigma _{1}(\omega )$, the rise and the peaks, to
interband transitions of Eu$_{1-x}$Ca$_{x}$B$_{6}$. Band structure
calculations show that in EuB$_{6}$, the CB width is about $3\sim 4$~eV and
the band bottom is formed at the X point of the Brillouin zone (BZ). The VB
top is located at the same point. Most of the calculations show that, in EuB$%
_{6}$, the two bands overlap by small amount to form a semimetallic state.
\cite{massidda,kunes} An optical excitation from the VB to the CB will then
occur over a broad frequency from $\omega =0$ to the maximal interband
energy. For instance, according to Massidda \textit{et al.}, the 2.75~eV
peak corresponds to the VB-CB excitation at the $\Gamma $ point.\cite
{massidda} The blue shift of $\sigma _{1}(\omega )$ which occurs nearly
uniformly over the wide frequency range implies that with Ca-doping, the
VB-CB distance increases throughout the BZ.

In EuB$_{6}$, the VB is formed by the B $2p$ level, while the CB is derived
from the hybridization of the cation $d$ (Eu $4d$) and B $2p$ levels.\cite
{tromp,massidda,rodriguez,hasegawa,kunes} As Ca is introduced, the lattice
constant of Eu$_{1-x}$Ca$_{x}$B$_{6}$ monotonically decreases.\cite{rhyee2}
Also, within the unit cell, the cation-boron distance decreases and as a
result, the hybridization will change. In addition, Eu$^{+2}$ has a larger
ionic radius than Ca$^{+2}$ (1.12~\AA\ vs 1.0~\AA ). These structural and
chemical changes seem to contribute to the observed band shift. Further, Kune%
{\v{s}} $et$ $al.$ showed that in EuB$_{6}$, the Eu 4$f$ state,
through hybridization with B 2$p$ state, has significant
consequences in the band structure, particularly at X point of
BZ.\cite{kunes} It will be interesting to perform a complete band
structure calculation of Eu$_{1-x}$Ca$_{x}$B$_{6}$ to see whether
the observed band shift is reproduced, and also to find which effect
plays the major role of it.\cite{com1} As for the 1.5 eV peak, Caimi
\textit{et al.} assigned it to the intra-atomic Eu $4f\rightarrow
5d$ transitions.\cite{caimi,vonlanthen} This is consistent with the
absence of the peak in CaB$_{6}$.

In Fig.~\ref{fig:sigma}(b), we show the low-frequency part of the spectrum.
The sharp peaks at $\omega \approx 146$ and 858~cm$^{-1}$ represent
infrared-active phonons. The rapid increase of $\sigma _{1}(\omega )$ at
low-frequency with decreasing $\omega $ represents the metallic response of
free carriers. This intraband conductivity decreases with $x$ and is
suppressed to a negligible amount for $x\geq 0.35$, indicating that the
metallic carriers disappear. We estimate the spectral weight of $\sigma
_{1}(\omega )$ in terms of the plasma frequency $\omega _{p}$ using the
relation $\omega _{p}^{2}=\frac{120}{\pi }\int_{0}^{\omega _{c}}\sigma
_{1}(\omega )d\omega $. Here the cut-off frequency $\omega _{c}$ for the
integration was taken as $2000$ cm$^{-1}$. The phonon contribution was
subtracted from the sum. The result is shown in Fig.~\ref{fig:data} which
will be discussed later. Additionally, we note that the intraband $\sigma
_{1}(\omega )$ does not follow the conventional Drude form. Perucchi \textit{%
et al.} analyzed it as a sum of several Drude components.\cite{perucchi}

In EuB$_{6}$, the metallic carrier creation is attributed either to an
extrinsic origin such as a boron vacancy\cite{denlinger} or to the intrinsic
semimetallic CB-VB overlap.\cite{massidda} Degiorgi \textit{et al.} measured
reflectivity of an independent EuB$_{6}$ sample.\cite{degiorgi} The plasma
edge in that work coincides with ours, suggesting that the carrier density
is same. In the present work, the plasma frequency changes systematically in
Ca-doped samples. It is less likely that these results come from
uncontrolled random vacancies. As for the latter scenario, which seems more
plausible at this point, the carrier density depends on the amount of the
band overlap at X-point. The $x$-dependence of $\omega _{p}$ suggests that
the overlap decreases with Ca-doping. For $x\geq 0.35$, it is inferred that
CB and VB are separated.

%
% Figure 3
%
\begin{figure}[tbp]
\vspace*{-0.5cm}\centerline{%
\includegraphics[width=2.6in,angle=-90]{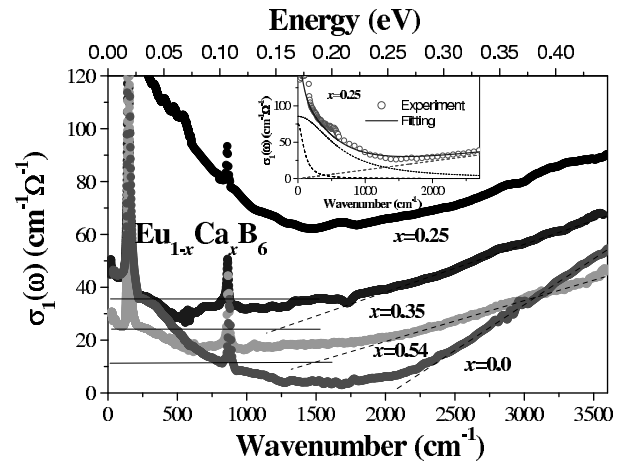}}
%\centerline{\includegraphics[width=4.5in,angle=-90]{fig3.eps}}%
\vspace*{0.0cm}%
%\centering\epsfig{file=fig3.eps,width=4cm,angle=-90}
%
\caption{Otical conductivity of Eu$_{1-x}$Ca$_{x}$B$_{6}$ below 0.5
eV. The curves are displaced vertically. The solid horizontal lines
show the displacements. The dashed lines are used to determine the
absorption edge of
the interband transition. (Inset): A decomposition of $\protect\sigma _{1}(%
\protect\omega )$ of $x=0.25$ into intraband and interband
contributions. We
use two Drude conductivities to fit the intraband $\protect\sigma _{1}(%
\protect\omega )$. The phonon peak was substracted from the data.}
\label{fig:mir}
\end{figure}

To test this picture more directly, let us give a closer look at the onset
of the interband conductivity which corresponds to the VB-CB\ excitation at
X point. In Fig.~\ref{fig:mir} we show $\sigma _{1}(\omega )$ in the
mid-infrared region for four samples, one below $x_{c}$ ($x=0.25)$ and the
rest at $x\geq x_{c}$. For a clearer presentation, the curves have been
displaced vertically. The interband $\sigma _{1}(\omega )$ appears as a
linear rise with increasing $\omega $ on the high energy side of the
spectra. For the samples with $x\geq 0.35$, \ we extrapolate along the
linear region to find the onset energy $\Delta $, as shown by the dashed
lines. At $x=1$ (CaB$_{6}$), $\Delta $ is about 0.25~eV. The Drude feature
in the far-infrared region represents perhaps residual impurity-induced
carriers. At $x=0.54$ and $0.35,$ $\Delta $ is smaller. For these two
samples, $\Delta $ determination is somewhat uncertain due to the remnant $%
\sigma _{1}(\omega )$ at $\omega <\Delta $.\cite{com2} At $x=0.25,$ the
intraband $\sigma _{1}(\omega )$ is strong. Various model functions fit to
the intraband $\sigma _{1}(\omega )$ were evaluated, but depending on the
fitting details, $\Delta $ varied from 0 to as large as 0.2~eV. The inset
illustrates a fit with $\Delta =0$ which corresponds to the case where the
bands overlap. The large uncertainty of $\Delta $ makes it difficult to
determine the X-point band state for $x<0.35$.

%
% Figure 4
%
\begin{figure}[tbp]
%
%
% manuscript
%
\vspace*{0.3cm}\centerline{\includegraphics[width=2.3in]{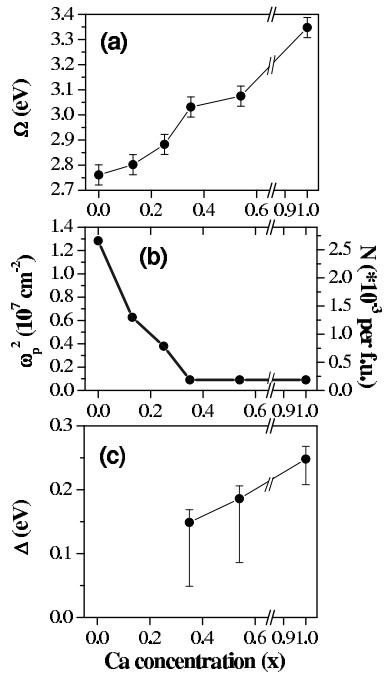}}
%\centerline{\includegraphics[width=3.8in]{fig4.eps}}%
\vspace*{0.0cm}%
%\centering\epsfig{file=fig4.eps,width=4cm,angle=0}
%
\caption{The changes of the optical features of
Eu$_{1-x}$Ca$_{x}$B$_{6}$
with $x$. (a) Position of the 2.7~eV peak. (b) Squared plasma frequency $%
\protect\omega _{p}^{2}$ (c) Absorption onset of the interband
transition. In (b), $\protect\omega _{p}^{2}$ is converted into
carrier density n (right axis) from $\protect\omega _{p}^{2}=4\pi
ne^2/m^*$ using $m^*$=0.25 [ref.18 and 19]. Error bars of the values
are indicated together.} \label{fig:data}
\end{figure}

In Fig.~\ref{fig:data}, we summarize the observed optical changes of
the 2.7~eV peak position $\Omega $, the squared plasma frequency
$\omega _{p}^{2}$ and the interband onset frequency $\Delta $. The
$\Omega $ in CaB$_{6}$ (=3.35 eV) is about 0.6 eV higher than in
EuB$_{6}$. Meanwhile, the band overlap in EuB$_{6}$ estimated from
the band calculation is about 0.3~eV.\cite{kunes} If the bands move
rigidly throughout the BZ, the X-point separation at $x=1$ will be
0.3~eV. This is close to the observed $\Delta =0.25$~eV of
CaB$_{6}$. Also the bands will fall apart ($\Delta =0)$ when the
shift is 0.3~eV. This occurs, from Fig.4(a), at $x$ $=0.35$, which
agrees with the suppression of $\omega _{p}^{2}.$ However, $\Delta $
at this composition ($=$ 0.15 eV) is rather significant. It may
indicate that the gap opens quickly, i.e, between $x$ $=0.25$ and
0.35, the band shift at the X-point occurs faster than the other
part of $x.$ Interestingly, the change of $\Omega $ shows a rapid
shift for this $x$ range. However, due to the uncertainty of $\Delta
$ at $x$ $=0.35$, it is difficult to draw the exact $x $-dependence
of the gap opening. Neverthless, the overall behaviors of the three
optical conductivity features show consistently that the electronic
structure transition occurs from the semimetallic EuB$_{6}$ to the
insulating CaB$_{6}$ through the band shift. EuB$_{6}$ exhibits a
ferromagnetic transition at T$_{c}=15$~K. It was found that $T_{c}$
decreases with Ca-substitution and then, interestingly, disappears at $x_{c}$%
.. This correlation suggests that the ferromagnetism is coupled with the
charge carrier.\cite{rhyee} A quantitative analysis of this relationship
will be discussed in an independent paper.\cite{jhkim}

%
% Summary
%
In summary, we have performed reflectivity and ellipsometry measurements on
the isovalent hexaboride series compounds Eu$_{1-x}$Ca$_{x}$B$_{6}$ ($0\leq
x\leq 1$) over wide energy range of 2.5 meV - 6 eV and found that the
optical conductivity exhibits systematic evolutions from EuB$_{6}$ to CaB$%
_{6}$. The interband $\sigma _{1}(\omega )$ including the two peaks
at 2.7 eV and 3.7 eV shifts continuously to higher energy as $x$
increases. This shows that, along with the cation substitution, the
valence band and the conduction band move away from each other
throughout the Brillouine zone and the VB to CB energy distance
increases with the Ca-content. As for the the low-frequency
intraband $\sigma _{1}(\omega )$, the Drude spectral weight
decreases continuously with $x$ and suppressed for $x>0.35.$ In
EuB6, we adopted the semimetallic band structure calculation results
and showed that the observed behaviors of the low frquency $\sigma
_{1}(\omega )$ are most readily explained in terms of the band shift
at the X point of BZ. The carrier density decreases due to the
decrease of the CB-VB overlap. At higher $x,$ the two bands are
separated to open a band gap, which is consistent with the optical
absorption onset at finite frequency, 0.25 eV in CaB$_{6}.$ As we
mentioned in the introduction, extensive efforts have been made to
calculate the band structure of EuB$_{6}$ and CaB$_{6}.$ The
observations of the present work, i.e, the systematic evolution of
the
electronic structure along with the cation substitution in Eu$_{1-x}$Ca$_{x}$%
B$_{6}$ provide a strong constraint on the band structure calculations and
should guide future works toward the complete understanding of the
hexaboride compounds.

This work was supported by the KRF Grant No. 2002-070-C00032 and the KOSEF
through CSCMR. Work at Brookhaven National Laboatory was supported by the
DOE under Contract No. DE-AC02-98CH10886.

%
%%%%%%%%%%%%%%%%%%%%%%%%%%%%%%%%%%%%%%%%%%%%%%%%%%%%%%%%%%%%%%%%%%%%%%%%%%%%%%
%
% References
%

\bibliography{hexaboride}

\begin{thebibliography}{40}
\bibitem[{com({\natexlab{a}})}]{com1}
\bibinfo{note}{In fact, in Ref.~\onlinecite{massidda}, the authors calculated
  the band structure of EuB$_{6}$ and CaB$_{6}$ using the LDA
  method.\cite{massidda} The result shows that the VB-CB distance at $\Gamma $
  point is smaller in CaB$_{6}$ by about 0.5 eV, contrary to the observation.}
\bibitem[{com({\natexlab{b}})}]{com2}
\bibinfo{note}{We have confirmed that $\sigma _{1}(\omega )$ in this energy
  range is insensitive to the low frequency extrapolation of the Kramers-Kronig
  transformation}.
\end{thebibliography}

\end{document}